\documentclass[aps,prl,reprint,groupedaddress,nofootinbib]{revtex4-1}

\usepackage{graphicx}
\usepackage{longtable}

\usepackage{relsize}
\def\Babar{{\mbox{\slshape B\kern-0.1em{\smaller A}\kern-0.1em B\kern-0.1em{\smaller A\kern-0.2em R}}}}

 \usepackage{amssymb} \usepackage{amsmath}
\usepackage{color}
\usepackage{url}
\usepackage[bookmarks, breaklinks, colorlinks,urlcolor=black, citecolor=red, 
linkcolor=blue]{hyperref}

\def\g {\gamma}

\def\bar {\overline}

\def\bea {\begin{eqnarray}}
\def\eea {\end{eqnarray}}

\def\rd {R(D)}
\def\rdst {R(D^*)}
\def\rdrdst {R(D^{(*)})}

\def\rk {R_K}
\def\rkst {R_{K^*}}
\def\rkrkst {R_{K^{(*)}}}

\def\st {\sin\theta}
\def\ct {\cos\theta}

\def\g{\gamma}

\def\beq{\begin{equation}}
\def\eeq{\end{equation}}
\def\barr{\begin{array}}
\def\earr{\end{array}}
\def\dis{\displaystyle}

\def\gev{\ensuremath{\mathrm{Ge\kern -0.1em V}}}
\def\tev{\ensuremath{\mathrm{Te\kern -0.1em V}}}
\def\Cn{\ensuremath{C^\mathrm{NP}\,}}
\def\Cten{\ensuremath{C_{10}^\mathrm{NP}\,}}
\def\Cnu{\ensuremath{C_\nu^\mathrm{NP}\,}}
\def\Cnine{\ensuremath{C_{9}^\mathrm{NP}\,}}
\usepackage{cancel}
 \usepackage[normalem]{ulem}
\usepackage{color}

\def\lapp{\mathrel{\rlap{\raise.5ex\hbox{$<$}}
                    {\lower.5ex\hbox{$\sim$}}}}
\def\gapp{\mathrel{\rlap{\raise.5ex\hbox{$>$}}
                    {\lower.5ex\hbox{$\sim$}}}}
\begin{document}

\title{Minimal unified resolution to $\boldmath {R_{K^{(*)}}}$ and $\boldmath {R(D^{(*)})}$ anomalies with lepton mixing}

\author{Debajyoti Choudhury$^1$, Anirban Kundu$^2$, Rusa Mandal$^3$ and 
  Rahul Sinha$^3$}
\affiliation{$^1$Department of Physics and Astrophysics, University of Delhi, Delhi 110007, India\\
$^2$Department of Physics, University of Calcutta,
92 Acharya Prafulla Chandra Road, Kolkata 700009, India\\
$^3$Institute of Mathematical Sciences, HBNI, Taramani, Chennai 600113, India}

%\date{\today}

\begin{abstract}
It is a challenging task to explain, in terms of a simple and
compelling new physics scenario, the intriguing discrepancies between
the standard model expectations and the data for the neutral-current
observables $R_K$ and $R_{K^*}$, as well as the charged-current
observables $R(D)$ and $R(D^*)$.  We show that this can be achieved in
an effective theory with only two unknown parameters.  In addition,
this class of models predicts some interesting signatures in the
context of both $B$ decays as well as high-energy collisions.
\end{abstract}
\pacs{13.20.He, 14.40.Nd, 11.30.Fs}
%\keywords{}
\maketitle

%%%%%%%%%%%%%%%%%%%%%%%%%%%%%%%%%%%%%%%%%%%%%%%%%%%%%%%%%%%%%

{\em Introduction and the data} -- Several recent hints of
discrepancies in a few charged- as well as neutral-current
semileptonic decays of $B$-mesons have intrigued the community. Unlike
the case for fully hadronic decay modes that suffer from large (and,
in some cases, not-so-well understood) strong interaction corrections, the
theoretical uncertainties in semileptonic decays are much better
controlled.  Even these uncertainties are removed to a great extent in
ratios of similar observables.  While, individually, none of the
observables, militate against the standard model (SM), viewed
together, they strongly suggest that some new physics (NP) is lurking
around the corner~\cite{globalfit,Bhattacharya:2014wla}.  The pattern also argues convincingly for the violation
of lepton-flavor universality.

With the ratios of partial widths being particularly clean probes of
physics beyond the SM, on account of the cancellation of the leading
uncertainties, let us focus on $R(D)$ and $R(D^*)$ defined as
%%%%%%%%
\beq
\label{eq:RD}
\rdrdst \equiv \frac{ {\rm BR}(B\to D^{(*)}\tau\nu)}{ {\rm BR}(B\to D^{(*)}\ell \nu)}\,,~~\ell \in \{e,\mu\}
\eeq
%with $\ell=e$ or $\mu$, 
and analogous ratios for the neutral-current sector
\beq
\label{eq:RK}
\rkrkst \equiv \frac{ {\rm BR}(B\to K^{(*)} \mu\mu)}{ {\rm BR}(B\to K^{(*)}e e)}\,.
\eeq

With the major source of uncertainty in the individual modes being the
form factors, they largely cancel out\footnote{The cancellation 
works best for relatively large momentum transfers (where the leptonic 
mass effects are negligible), the region with the best data.} in ratios like 
$\rdrdst$ or $\rkrkst$, and
the SM estimates for these ratios are rather robust. 
Several measurements of $\rd$ and $\rdst$ by the \Babar~\cite{Lees:2013uzd},
Belle~\cite{Huschle:2015rga,Abdesselam:2016cgx}, and 
LHCb~\cite{Aaij:2015yra,LHCb_rdst_3prong} Collaborations indicated an upward deviation from the 
SM expectations. Combining the individual results,
 namely, $R(D) = 0.407\pm 0.039\pm 0.024$ 
  and $R(D^*)= 0.304\pm 0.019\pm 0.029$,
   the discrepancies are at $\sim 2.3\sigma$ and $\sim 3.4 \sigma$
  respectively. On the inclusion of the correlation between the data, the
  combined significance is at the $\sim 4.1 \sigma$ level~\cite{hfag} from the SM predictions~\cite{Fajfer:2012vx}. 

The data on $\rk$ and $\rkst$, on the other hand, lie systematically 
below the SM expectations \cite{bifani,1406.6482}:
%%%%%
\beq
\label{data:RK}
\barr{rclcl}
\rk &=& \dis 0.745^{+0.090}_{-0.074}\pm 0.036 & \, \, &\dis q^2 \in [1:6]\, {\rm GeV}^2\,,\\[2ex]
\rkst^{\rm \,low} &=&\dis {0.660}^{+0.110}_{-0.070} \pm 0.024 &  &\dis  q^2 \in [0.045:1.1]\, {\rm GeV}^2\,,\\[2ex]
\rkst^{\rm \,cntr} &=& \dis 0.685 ^{+0.113}_{-0.069} \pm 0.047  &  &
        q^2 \in [1.1:6]\, {\rm GeV}^2\,.
\earr
\eeq
%%%
For both $\rk$ and $\rkst^{\rm \,cntr}$, the SM predictions are
virtually indistinguishable from unity~\cite{sm-pred}, whereas for
$\rkst^{\rm \,low}$ it is $\sim$ 0.9 (owing to a finite $m_\mu$).  
Except for $\rkst^{\rm \,low}$, the theoretical uncertainties have been 
subsumed in the experimental ones.
 Thus the
measurements of $\rk$, $\rkst^{\rm \,low}$ and $\rkst^{\rm \,cntr}$,
respectively, correspond to $2.6\sigma$, $2.1\sigma$ and $2.4\sigma$
shortfalls from the SM expectations.

For the $K^{*}$ mode, a discrepancy is visible not only in the ratios
of binned differential distribution for muon and electron modes but
also in some angular distributions, like the celebrated
$P'_5$~\cite{P5`-def} anomaly for the decay $B\to K^* \mu\mu$
\cite{LHCb:2015dla}, at more than $3\sigma$.  Restricting ourselves to 
only the low and medium-$q^2$ region, namely, $q^2\le
6\,\gev^2$ (as the high-$q^2$ region can be
affected by a different kind of physics~\cite{Ciuchini:2015qxb}),
we do not include this anomaly in our analysis. However,
we see later that our fitted Wilson coefficients can explain this discrepancy as pointed out in global 
fits~\cite{globalfit}.

A similar suppression (at a level of approximately $3\sigma$) is seen in
the observable $\Phi \equiv d {\rm BR}(B_s\to\phi\mu\mu)/ d
m_{\mu\mu}^2$ in the analogous bin ($m_{\mu\mu}^2\in[1:6]\,{\rm
GeV}^2$) ~\cite{Aaij:2015esa,Altmannshofer:2014rta,Straub:2015ica},
namely,
%%%%
\beq
\label{data:phimumu}
\Phi
= \left\{ \hskip-5pt
    \barr{lcl}
    \dis \left(2.58^{+0.33}_{-0.31}\pm 0.08\pm 0.19\right) \times 10^{-8}~{\rm GeV}^{-2} &  & ({\rm exp.}) \\[1.ex]
    \dis \left(4.81 \pm 0.56\right) \times 10^{-8}~{\rm GeV}^{-2} & & ({\rm SM})\,. 
  \earr
  \right.
\eeq
With low theoretical error, this bin is virtually the same as that for
$\rk$ and $\rkst^{\rm \,cntr}$.  This suggests strongly that the
discrepancies in the latter have been caused by a depletion of the $b
\to s \mu \mu$ channel, rather than an enhancement in $b \to s e
e$, a surmise further vindicated by the  $P'_5$ anomaly.
Note that $P'_5$ is dominated by the 
vector operator ${\cal O}_{9}$, while the two-body decay $B_s\to\mu\mu$ 
is controlled by the axial vector operator ${\cal O}_{10}$, both of them defined later.

With possible
corrections from large $\Delta\Gamma_s$, as well as next-to-leading-order (NLO) electroweak
and next-to-next-to-leading-order QCD corrections calculated, the SM prediction is quite robust
with only small uncertainties accruing from the Cabibbo-Kobayashi-Mashkawa (CKM) matrix elements
and the decay constant of $B_s$.  The LHCb measurement at a
significance of $7.8\sigma$ \cite{Bsmumu,BsmumuSM} shows an excellent
agreement between the data and the SM:
%%%
\beq
\label{data:Bsmumu}
{\rm BR}(B_s \to \mu \mu)=
\left\{\hskip-5pt
\barr{lcl}
\dis \left(3.0\pm 0.6^{+0.3}_{-0.2}\right)\times {10}^{-9} &\hskip-5pt & ({\rm exp.}), 
   \\[2ex]
 \dis \left(3.65 \pm 0.23\right) \times 10^{-9} &\hskip-5pt & ({\rm SM})\,,
\earr
\right.
\eeq
%%%
and hence puts very strong constraints on NP models, in particular on
those incorporating (pseudo)scalar or axial-vector
currents~\cite{Fleischer:2014jaa}. However, note that the central
value can accommodate a $\sim 20\%$ suppression.
Thus, one is naturally led to models that preferentially alter ${\cal O}_{9}$ rather than 
${\cal O}_{10}$.

Similarly, neither the radiative decay $B\to X_s\gamma$ nor the mass
difference $\Delta M_s$ and mixing phase $\phi_s$ measurements for the
$B_s$ system show any appreciable discrepancy with the SM
expectations.  The pattern of deviations is thus a complicated one and, naively at least, 
does not appear to show a definite direction towards any well-motivated NP model.
%seemingly contradictory. 
Consequently, most efforts at explaining the anomalies consider only a
subset, either $\rk$ and/or $\rdrdst$ data~\cite{oldlit,Altmannshofer:2008dz}, or $\rkrkst$ and $b\to s \ell \ell$ data~\cite{rknew}.
Those that do attempt a more complete treatment either invoke
very complicated models, or result in fits that are not very good.  In
addition, they are liable to result in other unacceptable
phenomenological consequences. Analyses within specific models, like
leptoquarks, are available in the literature~\cite{crivellin-march17}.

In view of this, we adopt a very phenomenological approach, rather than
advocate a particular model.  Assuming an effective Lagrangian, with
the minimal number of new parameters, in the guise of the unknown
Wilson coefficients (WCs), we seek the best fit. While not an entirely
new idea, our analysis takes into account not only the anomalous
channels but also the existing limits on several other channels; as we
will show, they provide the tightest constraints on the parameter
space. This approach hopefully will pave the way to unravelling the as
yet unknown flavor dynamics.

%-----------------------------------
{\em Models} -- Within the SM, the $ b\to c \tau \bar{\nu}_\tau$ transition
proceeds through a tree-level $W$ exchange. If the NP
adds coherently to the SM, one can write the effective Hamiltonian as
\beq
\label{eq:Hcnutau}
{\cal H}^{\rm eff} = \frac{4G_F}{\sqrt{2}} V_{cb} \left(1+\Cn \right) \, 
   \left[(c,b) (\tau,\nu_\tau)\right]\,,
\eeq
where the NP contribution is parametrized by $\Cn$  vanishes in the SM
limit and we have introduced the shorthand notation
%\beq
$(x,y) \equiv \bar x_L \g^\mu y_L \ \forall \ x,y\,.$
%\eeq
To explain the data, one thus needs either small positive or
large negative values of $\Cn$.  

The flavor-changing neutral-current decays $B\to K^{(*)} \mu \mu$ and  $\phi \mu \mu$ are 
occasioned by the $ b\to s \mu \mu$ transition proceeding, within the SM, 
primarily 
through a combination of the penguin and the box diagrams
(driven, essentially, by the top quark).  
Parametrizing the ensuing effective Hamiltonian as
\bea
{\cal H}^{\rm eff} = \frac{-4G_F}{\sqrt{2}} \, V_{tb} \, V_{ts}^*\, 
              \sum_{i}C_i(\mu) \mathcal{O}_i(\mu)\,,
\eea
where the relevant operators are 
%%%%%%%%%%%%%%%%%%%%
\beq
\barr{rcl}
\dis \mathcal{O}_7 & = & \dis (\alpha_{\rm em}(m_b) \, m_b / 4 \pi) \, 
  \left(\bar{s} \sigma_{\mu\nu} P_R b\right) 
   F^{\mu \nu} \,,  \\
\dis \mathcal{O}_9 & = & \dis (\alpha_{\rm em}(m_b) / 4 \pi) \, 
    \left(\bar{s} \g_\mu P_L b\right) \left(\bar{\mu} \g^\mu \mu\right) \,,\\ 
\dis \mathcal{O}_{10}& = & \dis  (\alpha_{\rm em}(m_b) / 4 \pi) \, 
     \left(\bar{s} \g_\mu P_L b\right) \left(\bar{\mu} \g^\mu\g_5 \mu\right)\, .
\earr
\eeq
The WCs, matched with the full theory at $m_W$ and then run down to
$m_b$ at the next-to-next-to-leading logarithmic accuracy
\cite{Altmannshofer:2008dz}, are given in the SM as $C_7 = -0.304
\,,\ \ C_9 = 4.211 \,{\rm ~and~} \ C_{10}= -4.103 \,.$ The differential 
widths for the $B\to K^{(*)} \mu \mu$ decay are obtained in terms of 
algebraic functions of these. NP contributions to ${\cal H}^{\rm eff}$ 
can be parametrized by $C_i \to C_i + C_i^{\rm NP}$. 

Similarly, the $ b\to s \nu \bar{\nu}$ 
transition (which governs the $B\to K^{(*)} \nu \bar{\nu}$ decays) 
proceeds through the $Z$ penguins and box diagrams. 
Unless right-handed neutrino fields are introduced, the 
low energy effective Hamiltonian can be parametrized
by~\cite{MET_SM}
%%%%%
\bea
{\cal H}^{\rm eff} = \frac{2G_F}{\sqrt{2}} \, V_{tb} \, V_{ts}^*\,
   \frac{\alpha_{\rm  em}}{\pi} \, C_L^{\rm SM}  \left(1+ \Cnu\right)
(s,b)(\nu,\nu),
\eea
%%%%
where \Cnu denotes the NP contribution. Including the NLO QCD correction
and the two-loop electroweak contribution, the SM WC is given by
$C_L^{\rm SM}=-X_t/s_w^2$ where the Inami-Lim function $X_t= 1.469 \pm
0.017$~\cite{MET_SM_old,MET_SM}.

While it may seem trivial to write down extra four-fermi operators that
would produce just the right contributions, care must be taken to see
that this does not introduce unwelcome consequences. For one, a large
enhancement of $C_{10}$ could lead an unacceptably large ${\rm BR}(B_s
\to \mu \mu)$, with $ \mathcal{O}_{10}$ being the leading contributor
to this decay. Similarly, the said four-fermi operators need to be
invariant under the SM gauge group (assuming that the NP appears only
above the electroweak scale). A non-zero \Cn (see Eq.\
(\ref{eq:Hcnutau})) would, potentially, lead to an analogue of
$C_{10}^{\rm NP}$ for the tau-channel.  This, in turn, would lead to
an enhancement of $B_s \to \tau \tau$, where the chirality suppression
is less operative than in the muonic case.  Indeed, the LHCb
Collaboration~\cite{Aaij:2017xqt} has obtained a 95\% C.L. upper limit
of $6.8 \times 10^{-3}$ on the branching fraction for this
mode\footnote{It should be noted, though, that this analysis does not
actually reconstruct the $\tau$s, but employs neural networks.  Hence,
it is possible that future measurements would point to a value higher
than the limits quoted.}, with the SM value being $(7.73\pm
0.49)\times 10^{-7}$ \cite{BsmumuSM}. Similarly, none of the three
operators $(b,s) \, (\nu_i, \nu_i)$ may receive large corrections lest
the SM expectations, namely~\cite{MET_SM}
%%%%
\beq
\label{SM:MET}
\barr{rcl}
\dis {\rm BR}(B^+ \to K^+ \nu \bar\nu)_{\rm SM} &\!\! =\!\! & \dis 
(3.98 \pm 0.43 \pm 0.19) 
\times 10^{-6} \ , 
\\[2ex]
\dis {\rm BR}(B^0 \to K^{*0} \nu \bar\nu)_{\rm SM} & \!\!=\!\! & \dis 
(9.19 \pm 0.86 \pm 0.50) \times 10^{-6},
\earr
\eeq
be augmented\footnote{Note that the neutrino flavors need not be
identical for the NP.}  to levels beyond the 90\% C.L upper bounds
(summed over all three neutrinos) as obtained by the Belle
Collaboration \cite{belle17}, {\em viz.}
\beq
\label{data:MET}
{\rm BR}(B\to K^{(*)}\nu\bar\nu) < 1.6 \, (2.7) \times 10^{-5}\, .
\eeq

%%%%%%%%%%%%%%%%%%%%%%%%%%%%%%%%%%%%%%%%%%%%%%%%%%

In view of the aforementioned constraints, we consider only a
combination of two four-fermi operators, characterized by a single WC
(assumed to be real to avoid new sources of $CP$ violation).  Since we
do not claim to obtain the ultraviolet completion thereof, we do not
speculate on the (flavor) symmetry that would have led to such a
structure, which could have arisen from a plethora of NP scenarios,
such as models of (gauged) flavor, leptoquarks (or, within the
supersymmetric paradigm, a breaking of $R$ parity) etc. To wit, we
propose a model involving two four-fermi operators in terms of the
second- and third-generation (weak-eigenstate) fields
\begin{align}
    \label{eq:H_NP}
\hspace*{-0.35cm}{\cal H}^{\rm NP}
&= A_1 \,(\bar Q_{2L} \g_\mu L_{3L})\,   (\bar L_{3L} \g^\mu Q_{3L}) \nonumber \\
&+ A_2 \,(\bar Q_{2L} \g_\mu Q_{3L}) \, (\bar \tau_R \g^\mu \tau_R)
\end{align}
where the overall Clebsch-Gordan coefficients have been subsumed 
and we demand $A_2 = A_1$. 

This operator, seemingly, contributes to
$\rdrdst$ but not to the other anomalous processes. 
This, though, is true only above the
electroweak scale. Below this scale, the Hamiltonian 
needs to be rediagonalized\footnote{With NP only modifying the 
Wilson coefficients of certain SM operators to a small extent,
the QCD corrections (as well as hadronic uncertainties) are analogous. 
Additional effects due to operator mixings are 
too small to be of any concern.}
In the quark sector, this is determined
by the quark masses and the small non-alignment due to $A_{1,2}$ can
be neglected. In the leptonic sector, though, the extreme smallness of the
neutrino masses implies that the nonuniversal term ${\cal H}^{\rm NP}$
plays a major role~\cite{ggl}. To this end, we
consider the simplest of field rotations for the left-handed leptons
from the unprimed (flavor) to the primed (mass) basis, namely
%%%%
\beq
\label{eq:rot}
\tau = \ct \, \tau' + \st \, \mu' \ ,
\qquad 
\nu_\tau = \ct \, \nu_\tau' + \st \, \nu_\mu'\,.
\eeq
%%%
This, immediately, generates a term with the potential to explain the
$b\to s \mu \mu$ anomalies. 
%However, no new amplitude is generated for
%$b \to s \bar \nu \nu$, thereby trivially satisfying the corresponding
%constraints.

%%%%%%%%%%%%%%%%%%%%%%%%%

{\em Results} --- The scenario  is, thus, characterized by two 
parameters, namely $A_1$ and $\sin\theta$. The best fit
values for these can be obtained by effecting a $\chi^2$-test defined
through
%%%
\bea
\label{eq:chisquare}
\chi^2= \sum_{i=1}^{7} \frac{\left(\mathcal{O}_i^{\rm exp}-\mathcal{O}_i^{\rm th} \right)^2}{\left(\Delta \mathcal{O}_i^{\rm exp}\right)^2+ \left(\Delta \mathcal{O}_i^{\rm th}\right)^2}
\eea
%%%
where $\mathcal{O}_i^{\rm exp}$ ($\mathcal{O}_i^{\rm th}$) denote the
experimental (theoretical) mean and $\Delta \mathcal{O}_i^{\rm exp}$
($\Delta \mathcal{O}_i^{\rm th}$) the corresponding $1\sigma$
uncertainty, with the theoretical values depending on the model
parameters.  We include a total of seven measurements for the
evaluation of $\chi^2$, namely, $\rd$, $\rdst$, $\rk$,
$\rkst^{\rm \,low}$, $\rkst^{\rm \,cntr}$, $\Phi$, and ${\rm
BR}(B_s\to\mu\mu)$ (while not affected by the NP interactions in
Eq.~\eqref{eq:H_NP}, this is relevant for the scenario considered later).
Only for the last two observables, do $\Delta\mathcal{O}_i^{\rm th}$
need to be considered explicitly , while, for the rest, they have been
subsumed within the experimental results. For our numerical analysis,
we use
%\[
$V_{cb}=0.0416 {\rm~and~} V_{tb}V_{ts}^*=-0.0409,$
%\]
and find, for the SM, $\chi^2_{\rm SM} \simeq 46$.  

Within the new model, the best fit corresponds to $\chi^2_{\rm min} \simeq 9$ 
(denoting a marked improvement) with the NP contributions being 
$C_9^{\rm NP}=-1.7$ and $C^{\rm NP}=-2.12$. In terms of the model parameters,
this corresponds to (note that there is a $\theta \to -\theta$ degeneracy)
%%%
\beq
A_1 (=A_2) = -2.92 \,\, \tev^{-2}\ , \qquad \sin\theta=\pm0.022\,,
\eeq
Even this low value of $\chi^2_{\rm min}$ is largely dominated by a
single measurement, namely, $\rkst^{\rm \,low}$. This is not
unexpected, as an agreement to this experimental value to better than
$1 \sigma$ is not possible if the NP contribution can be expressed
just as a modification of the SM WCs, rather than through the
introduction of a new and small dynamical scale (such a change 
could be tuned so as to manifest itself primarily only 
in the low-$q^2$ region, but is likely to have other ramifications).  Note that the small value of $\sin\theta$ can
only partially explain the atmospheric neutrino oscillation, while the
full explanation needs additional dynamics.
%%%%%%%%%%%%%%%%%%%
\begin{figure}[!ht]
	\begin{center}
\includegraphics[width=0.7\linewidth]{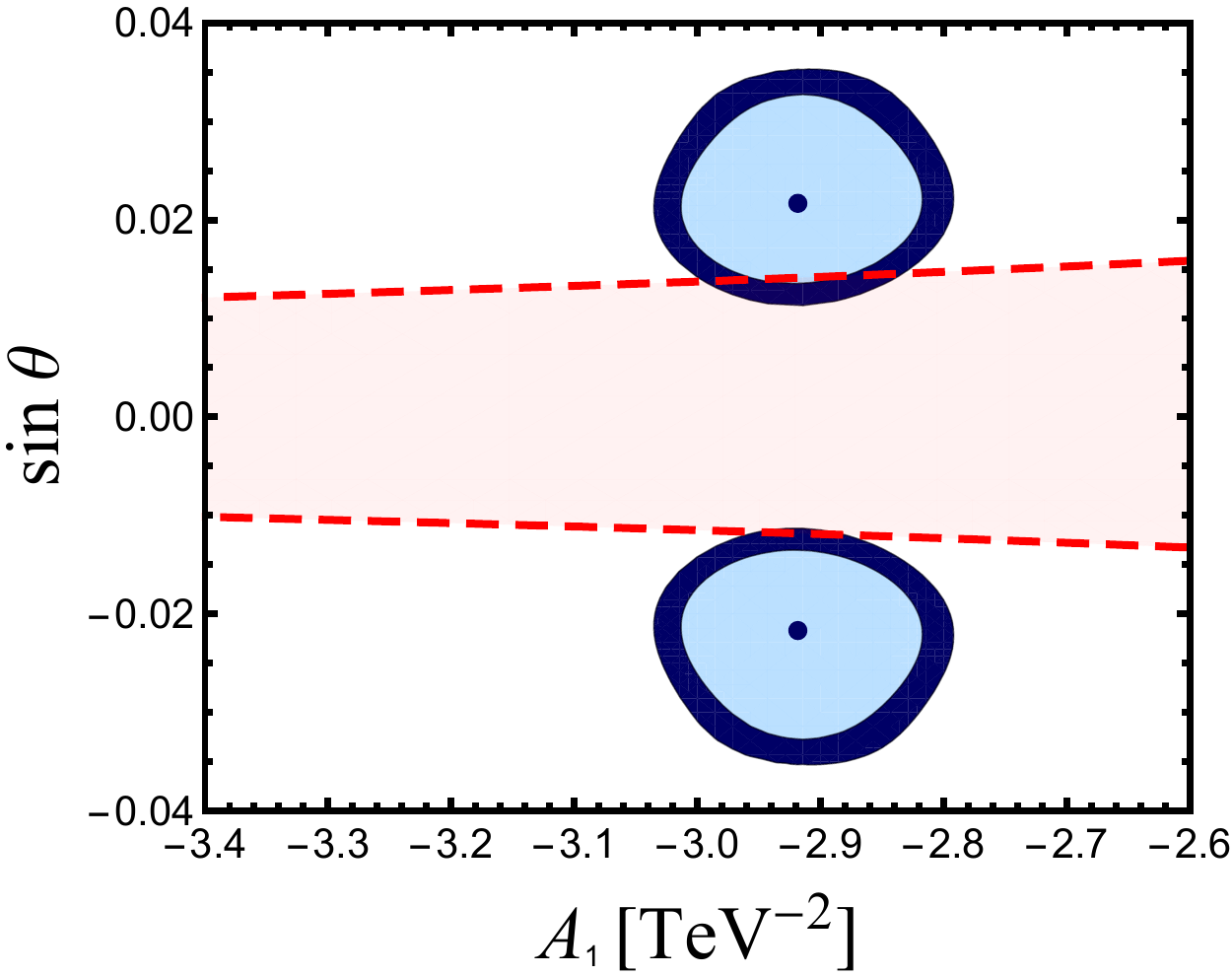}
\vskip -5pt
\caption{The light and dark blue regions denote $95\%$ and  $99\%$ C.L. 
bands, respectively, around the best-fit points. The red shaded region is allowed 
by bounds from BR$(B^+ \to K^+ \mu^- \tau^+)$.
}
	\label{fig:best_fit}
	\end{center}
\vskip -10pt
\end{figure}

More importantly, in effecting the field rotation of
Eq.\ (\ref{eq:rot}) in ${\cal H}^{\rm NP}$, we generate terms of the
form $(s, b)(\mu, \tau)$, leading to potential lepton-flavor
violating (LFV) decays. The current limits on the relevant ones are~\cite{pdg}
%%%%
\beq
\label{data:mutau}
{\rm BR}(B^+\to K^+\mu^\pm\tau^\mp) < 4.5 \, (2.8) \times 10^{-5}\,.
\eeq
%%%%
In Fig.~\ref{fig:best_fit}, we display the constraints from this
particular mode. While the best-fit point is summarily ruled out,
clearly solutions can be found if a slight worsening of the $\chi^2$
(to $\simeq 15$) is acceptable. This would still represent a much
better agreement than is possible within the SM. The corresponding
values of the observables are: $R_K=0.86$, $\rkst^{\rm \,cntr}=0.88$,
$\rkst^{\rm \,low}=0.90$, $R \left(D^{(*)}\right)= 1.25 \times R_{\rm SM}\left(D^{(*)}\right)$, and $\Phi=4.1 \times
{10}^{-8} \, \gev^{-2}$, representing quite a reasonable fit to all but
$\rkst^{\rm \,low}$. It should be noted here that the $\theta \to -
\theta$ degeneracy is broken by the LFV constraint, with $\theta > 0$
being slightly preferable.

\begin{figure}[!t]
	\begin{center}
\includegraphics[width=0.9\linewidth, height=0.65\linewidth]{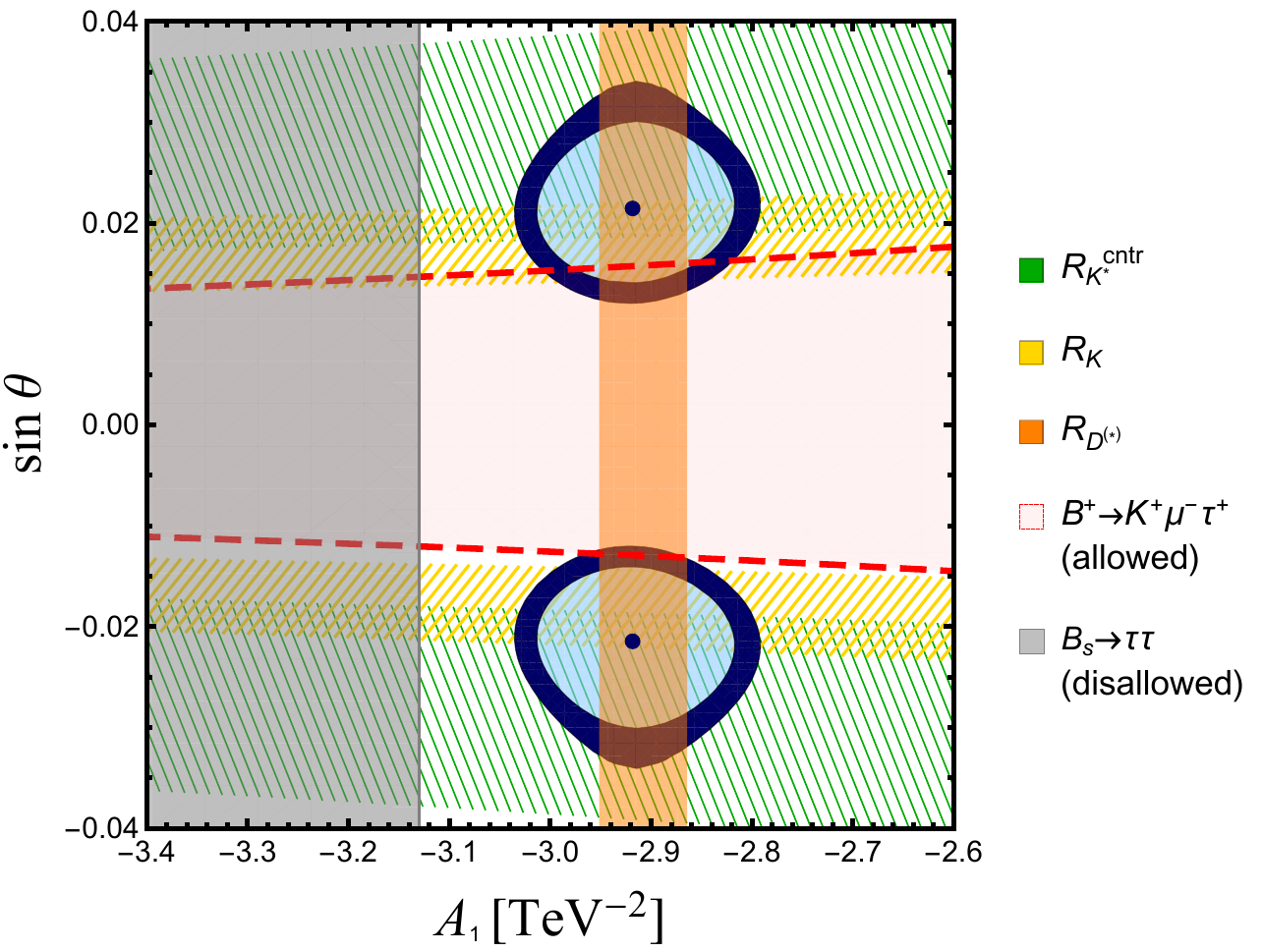}% {mod3aExt.pdf}
\vskip -5pt
\caption{The fit for $A_2 = 4 A_1/5$, with the bands around the best-fit 
points corresponding to 95\% and 99\% C.L. Also shown are the 
 $1 \sigma$ bands from $\rkrkst$ and $R(D)$, and the 95\% upper limits from 
$B_s \to \tau\tau$  and $B^+ \to K^+ \mu^-\tau^+$.
}
	\label{fig:best_fit2}
	\end{center}
\vskip -10pt
\end{figure}

Further improving the fit to $\rkrkst$ requires the introduction of a
small bit of $C_{10}^{\rm NP}$. Postponing the discussion of $B_s \to
\tau\tau$, this is most easily achieved if we choose to destroy, to a
small degree, the relation $A_2 = A_1$. As an illustrative example, we
consider $A_2= 4 A_1/5$. The consequent best fit values for $A_1$ and
$\st$ remain virtually the same but, now, $\chi^2_{\rm min}=7 $ with
NP contributions being $\Cnine=-1.51$, $\Cten=0.17$ and $C^{\rm
NP}=-2.12$. The result is depicted in Fig.~\ref{fig:best_fit2}. 
Once the LFV constraint is imposed, the observables at the overlap region are
$R_K\simeq0.80$, $\rkst^{\rm \,cntr}\simeq0.83$, $\rkst^{\rm \,low}\simeq0.88$,
$R \left(D^{(*)}\right) \simeq 1.24 \times R_{\rm SM}\left(D^{(*)}\right)$, and $\Phi \simeq3.8 \times {10}^{-8} \, \gev^{-2}$, showing marked improvement in the fit to all but $\rkst^{\rm \,low}$ and correspond to $\chi^2\simeq 10$. 
While the finite contribution to \Cten does enhance $B_s \to
\tau\tau$, the latter (gray shaded region in Fig.~\ref{fig:best_fit2}) 
does not have a major impact. It should be realized, though, that a
stronger breaking of the $A_2 = A_1$ relation would have led to a
better (worse) agreement with the LFV ($B_s \to \tau\tau$)
constraints.

It is interesting to speculate on the origin of this split between the
$A_i$. A naive explanation would be to attribute the difference to the
quantum numbers of the leptonic fields under an as yet unidentified
gauge symmetry, with the attendant anomaly cancellation being effected
by either invoking heavier fermionic fields or through other
means. Care must be taken, however, not to induce undesirable
phenomenology. An alternative is to attribute the
difference to quantum corrections,  although the aforementioned shift
is somewhat larger than that expected from a naive renormalization
group flow perspective, namely, $\sim (\alpha_{\rm wk.} / 4 \pi)
\ln(\Lambda_{\rm NP}^2 / m_b^2)$, where $\Lambda_{\rm NP} \sim $1 TeV
is the putative scale of NP. It should be noted here, though, that the
20\% shift is only illustrative and not really needed. Indeed, once
the electroweak symmetry is broken, the various pieces in ${\cal
H}^{\rm eff}$ suffer differing renormalization group flow down to the $m_b$ scale, and the
consequent breaking of the degeneracy is, putatively, of the right
magnitude to explain the remaining discrepancies.

It is worthwhile, at this stage, to explore the consequences of
introducing other operators in ${\cal H}^{\rm NP}$. While operators
constructed out of $SU(2)_L$-triplet currents (denoted by the subscript `3')
 such as $(\bar Q_{2L}
\g^\mu Q_{3L})_3 \, (\bar L_{3L} \g_\mu L_{3L})_3$, $(\bar Q_{2L}
\g^\mu L_{3L})_3 \, (\bar L_{3L} \g_\mu Q_{3L})_3$, etc., would also
have admitted solutions to the anomalies, they, typically, would also
result in unsuppressed $b \to s \nu \bar\nu$ transitions.
Circumventing the bounds would, then, require the introduction of
multiple operators and cancellations between them. We will discuss
such possibilities in detail in a subsequent paper.

This would, typically, still leave behind too large a rate for $B_s
\to \tau \tau$ (first reference of~\cite{crivellin-march17}) and, hence needs the further introduction of yet
another operator such as the second one in ${\cal H}^{\rm NP}$.  
Apart from enhancing $B_s \to \tau \tau$ ($B \to X_s \tau\tau$
and $\Lambda_b \to \Lambda \tau\tau$ are affected too, but bounds from
these sectors are not too serious), this would also affect the other
modes to varying degrees. Consequently, the best fit values will
change. Indeed a lower $\chi^2 \, (\simeq 5.4$) is achievable for
virtually the same $A_1$, but slightly smaller $|\sin\theta| \,
(\simeq 0.018)$.  Understandably, if both the $B_s \to \tau \tau$ bound 
as well that in Eq.\ (\ref{data:mutau}) are
to be satisfied, the $\chi^2$ can be reduced to at most $\simeq 11$.
Similarly, ${\rm BR}(B\to X_s\tau\tau)$, as well as ${\rm
BR}(\Lambda_b\to
\Lambda \tau\tau)$ will also be increased and should be 
close to observation at the LHCb.  However, processes like $b\to
s\gamma$ or $\tau\to\mu\gamma$ will remain under control, as we have
checked. Similarly, while we do not ``explain'' $(g-2)_\mu$, the agreement is 
marginally better than within the SM.
The new operators also generate, through renormalization group running,
operators involving four leptons \cite{paradisi}, and thus may lead to effects like $\tau\to 3\mu$. They are, 
however, well within control, mostly because of the small value of $\sin\theta$.

In summary, we have identified the minimal modification to the SM in
terms of an effective theory that can explain the anomalies in both
the charged- and the neutral-current decays of bottom mesons, a task
that has been challenging on account of the seemingly contradictory
requirements that the data demand. We circumvent this by postulating
just two four-fermi operators with WCs related by a
symmetry and taking advantage of the possibility of a small but
nontrivial rotation of the charged lepton fields that a
flavor-nonuniversal operator entails.  Taking all the data into
account, we find that with just two new parameters, the $\chi^2$ can
be reduced from 46 (in the SM) to below 15 while being consistent with
all other data. For the best fit point, most observables are
consistent within $\sim 1\sigma$, while $R_{K^*}$ and
BR$(B \to \phi \mu \mu)$ in the low-$q^2$ bins, are
consistent to only within $\sim 2\sigma$. 

The scale of new physics that such an explanation demands is a few
TeV at best, rendering searches at the LHC to be very interesting. An
even stronger preference is that at least one of $B\to K^{(*)}\mu\tau$
and $B_s \to \tau\tau$ should be close to discovery. A more precise
determination of the ratios that we have discussed in this Letter is,
therefore, of prime importance, as this can open the door to new
flavor dynamics and hence the world beyond the SM.

%%%%%%%%%%%%%%%%%%%%%%%%%%%%%%%%%%%%%%%%%%%%%%%%%%%%%%%%%%%%%
A.K. thanks the Science and Engineering Research Board (SERB), Government
of India, for a research grant. D.C. thanks the IMSc, Chennai for hospitality 
for part of the duration of the collaboration.

%\appendix
%\newpage
%\input{appendix.tex}

%%%%%%%%%%%%%%%%%%%%%%%%%%%%%%%%%%


\begin{thebibliography}{99}

%\cite{Mandal:2015bsa}
\bibitem{globalfit} 
%\cite{Altmannshofer:2013foa}
%\bibitem{Altmannshofer:2013foa} 
  W.~Altmannshofer and D.M.~Straub,
  %``New Physics in $B \to K^*\mu\mu$?,''
  Eur.\ Phys.\ J.\ C {\bf 73}, 2646 (2013);
%  doi:10.1140/epjc/s10052-013-2646-9
%  [arXiv:1308.1501 [hep-ph]].
  %%CITATION = doi:10.1140/epjc/s10052-013-2646-9;%%
  %202 citations counted in INSPIRE as of 26 Jun 2017
%\cite{Descotes-Genon:2013wba}
%\bibitem{Descotes-Genon:2013wba} 
  S.~Descotes-Genon, J.~Matias and J.~Virto,
  %``Understanding the $B\to K^*\mu^+\mu^-$ Anomaly,''
  Phys.\ Rev.\ D {\bf 88}, 074002 (2013);
%  doi:10.1103/PhysRevD.88.074002
%  [arXiv:1307.5683 [hep-ph]].
  %%CITATION = doi:10.1103/PhysRevD.88.074002;%%
  %282 citations counted in INSPIRE as of 14 Aug 2017

%\cite{Bhattacharya:2014wla}
\bibitem{Bhattacharya:2014wla} 
  B.~Bhattacharya {\em et al.}, %, A.~Datta, D.~London and S.~Shivashankara,
  %``Simultaneous Explanation of the $R_K$ and $R(D^{(*)})$ Puzzles,''
  Phys.\ Lett.\ B {\bf 742}, 370 (2015);
%  doi:10.1016/j.physletb.2015.02.011
%  [arXiv:1412.7164 [hep-ph]].
  %%CITATION = doi:10.1016/j.physletb.2015.02.011;%%
  %100 citations counted in INSPIRE as of 30 Jun 2017
%
R.~Mandal and R.~Sinha,
%``Implications from ${B\to K^*\ell^+\ell^-}$ observables using $3 \text{fb}^{-1}$ of LHCb data,''
Phys.\ Rev.\ D {\bf 95}, 014026 (2017);
%  doi:10.1103/PhysRevD.95.014026
%[arXiv:1506.04535 [hep-ph]];\\
%%CITATION = doi:10.1103/PhysRevD.95.014026;%%
%7 citations counted in INSPIRE as of 29 May 2017
%\cite{Karan:2016wvu}
%\bibitem{Karan:2016wvu} 
A.~Karan {\it et al.}, % R.~Mandal, A.~K.~Nayak, R.~Sinha and T.~E.~Browder,
  %``Signal of right-handed currents using $B\to K^*\ell^+\ell^-$ observables at the kinematic endpoint,''
  Phys.\ Rev.\ D {\bf 95}, 114006 (2017).
%  doi:10.1103/PhysRevD.95.114006
%  [arXiv:1603.04355 [hep-ph]].
  %%CITATION = doi:10.1103/PhysRevD.95.114006;%%
  %4 citations counted in INSPIRE as of 23 Jun 2017

\bibitem{Lees:2013uzd}
  J.P.~Lees {\it et al.} [BaBar Collab.],
  %``Measurement of an Excess of $\bar{B} \to D^{(*)}\tau^- \bar{\nu}_\tau$ Decays and Implications for Charged Higgs Bosons,''
  Phys.\ Rev.\ D {\bf 88}, 072012 (2013)
%  [arXiv:1303.0571 [hep-ex]].
  %%CITATION = doi:10.1103/PhysRevD.88.072012;%%

%\cite{Huschle:2015rga}
\bibitem{Huschle:2015rga}
  M.~Huschle {\it et al.} [Belle Collab.],
  %``Measurement of the branching ratio of $\bar{B} \to D^{(\ast)} \tau^- \bar{\nu}_\tau$ relative to $\bar{B} \to D^{(\ast)} \ell^- \bar{\nu}_\ell$ decays with hadronic tagging at Belle,''
  Phys.\ Rev.\ D {\bf 92}, 072014 (2015)
%  [arXiv:1507.03233 [hep-ex]].
  %%CITATION = doi:10.1103/PhysRevD.92.072014;%%
  %34 citations counted in INSPIRE as of 01 févr. 2016

\bibitem{Abdesselam:2016cgx}
  A.~Abdesselam {\it et al.} [Belle Collab.],
  %``Measurement of the branching ratio of $\bar{B}^0 \rightarrow D^{*+} \tau^- \bar{\nu}_{\tau}$ relative to $\bar{B}^0 \rightarrow D^{*+} \ell^- \bar{\nu}_{\ell}$ decays with a semileptonic tagging method,''
  arXiv:1603.06711 [hep-ex].
  %%CITATION = ARXIV:1603.06711;%%
  %4 citations counted in INSPIRE as of 07 May 2016

%\bibitem{Hirose:2016wfn} 
%  S.~Hirose {\it et al.} [Belle Collab.],
%  %``Measurement of the $\tau$ lepton polarization and $R(D^*)$ in the decay $\bar{B} \to D^* \tau^- \bar{\nu}_\tau$,''
%  arXiv:1612.00529 [hep-ex].
%  %%CITATION = ARXIV:1612.00529;%%
%  %1 citations counted in INSPIRE as of 08 Dec 2016  
  
%\cite{Aaij:2015yra}
\bibitem{Aaij:2015yra}
  R.~Aaij {\it et al.} [LHCb Collab.],
  %``Measurement of the ratio of branching fractions $\mathcal{B}(\bar{B}^0 \to D^{*+}\tau^{-}\bar{\nu}_{\tau})/\mathcal{B}(\bar{B}^0 \to D^{*+}\mu^{-}\bar{\nu}_{\mu})$,''
  Phys.\ Rev.\ Lett.\  {\bf 115}, 111803 (2015).
%  [Phys.\ Rev.\ Lett.\  {\bf 115}, no. 15, 159901 (2015)]
%  [arXiv:1506.08614 [hep-ex]].
  %%CITATION = doi:10.1103/PhysRevLett.115.159901, 10.1103/PhysRevLett.115.111803;%%
  %45 citations counted in INSPIRE as of 09 Feb 2016
\bibitem{LHCb_rdst_3prong}  
http://lhcb-public.web.cern.ch/lhcb-public/Welcome.html\#RDst2

 \bibitem{hfag}
  Y.~Amhis {\it et al.},
  %``Averages of $b$-hadron, $c$-hadron, and $\tau$-lepton properties as of summer 2016,''
  arXiv:1612.07233 [hep-ex], and the update at   http://www.slac.stanford.edu/xorg/hfag/semi/ \\   fpcp17/RDRDs.html.
    %%CITATION = ARXIV:1612.07233;%%  

%\cite{Fajfer:2012vx}
\bibitem{Fajfer:2012vx} 
  S.~Fajfer, J.F.~Kamenik and I.~Nisandzic,
  %``On the $B \to D^* \tau \bar \nu_{\tau}$ Sensitivity to New Physics,''
  Phys.\ Rev.\ D {\bf 85}, 094025 (2012);
%  doi:10.1103/PhysRevD.85.094025
%  [arXiv:1203.2654 [hep-ph]].
  %%CITATION = doi:10.1103/PhysRevD.85.094025;%%
  %264 citations counted in INSPIRE as of 14 Aug 2017
%\cite{Bigi:2016mdz}
%\bibitem{Bigi:2016mdz} 
  D.~Bigi and P.~Gambino,
  %``Revisiting $B\to D \ell \nu$,''
  Phys.\ Rev.\ D {\bf 94}, 094008 (2016).
%  doi:10.1103/PhysRevD.94.094008
%  [arXiv:1606.08030 [hep-ph]].
  %%CITATION = doi:10.1103/PhysRevD.94.094008;%%
  %27 citations counted in INSPIRE as of 14 Aug 2017
  
\bibitem{bifani}
R.~Aaij {\it et al.} [LHCb Collaboration],
  %``Test of lepton universality with $B^{0} \rightarrow K^{*0}\ell^{+}\ell^{-}$ decays,''
  arXiv:1705.05802 [hep-ex].
  %%CITATION = ARXIV:1705.05802;%%
  %S.~Bifani (on behalf of the LHCb Collab.), CERN seminar on April 18, 2017.

  \bibitem{1406.6482}
R.~Aaij {\it et al.} [LHCb Collab.],
%``Test of lepton universality using $B^{+}\rightarrow K^{+}\ell^{+}\ell^{-}$ decays,''
Phys.\ Rev.\ Lett.\  {\bf 113}, 151601 (2014).
%doi:10.1103/PhysRevLett.113.151601
%[arXiv:1406.6482 [hep-ex]].
%%CITATION = doi:10.1103/PhysRevLett.113.151601;%%  

  \bibitem{sm-pred}
G.~Hiller and F.~Kruger,
%``More model-independent analysis of $b \to s$ processes,''
Phys.\ Rev.\ D {\bf 69}, 074020 (2004);
%doi:10.1103/PhysRevD.69.074020
%[hep-ph/0310219]; \\
%%CITATION = doi:10.1103/PhysRevD.69.074020;%% 
M.~Bordone, G.~Isidori and A.~Pattori,
%``On the Standard Model predictions for $R_K$ and $R_{K^*}$,''
Eur.\ Phys.\ J.\ C {\bf 76}, 440 (2016);
%doi:10.1140/epjc/s10052-016-4274-7
%[arXiv:1605.07633 [hep-ph]]; \\
%\cite{DescotesGenon:2012zf}
\bibitem{P5`-def} 
  S.~Descotes-Genon {\it et al.}, % J.~Matias, M.~Ramon and J.~Virto,
  %``Implications from clean observables for the binned analysis of $B -> K*\mu^+\mu^-$ at large recoil,''
  JHEP {\bf 1301}, 048 (2013)
%  doi:10.1007/JHEP01(2013)048
%  [arXiv:1207.2753 [hep-ph]].
  %%CITATION = doi:10.1007/JHEP01(2013)048;%%
  %154 citations counted in INSPIRE as of 14 Aug 2017


%\cite{LHCb:2015dla}
\bibitem{LHCb:2015dla} 
R.~Aaij {\it et al.} [LHCb Collab.],
%``Angular analysis of the $B^{0} \to K^{*0} \mu^{+} \mu^{−}$ decay using 3 fb$^{−1}$ of integrated luminosity,''
JHEP {\bf 1602}, 104 (2016),
%  doi:10.1007/JHEP02(2016)104
%[arXiv:1512.04442 [hep-ex]].
%%CITATION = doi:10.1007/JHEP02(2016)104;%%
%3 citations counted in INSPIRE as of 01 Mar 2016

\bibitem{Ciuchini:2015qxb} 
M.~Ciuchini {\it et al.}, % M.~Fedele, E.~Franco, S.~Mishima, A.~Paul, L.~Silvestrini and M.~Valli,
%``$B\to K^* \ell^+ \ell^-$ decays at large recoil in the Standard Model: a theoretical reappraisal,''
JHEP {\bf 1606}, 116 (2016);
%    doi:10.1007/JHEP06(2016)116
%[arXiv:1512.07157 [hep-ph]]
%%CITATION = doi:10.1007/JHEP06(2016)116;%%
%21 citations counted in INSPIRE as of 20 Sep 2016

%\cite{Aaij:2015esa}
\bibitem{Aaij:2015esa} 
R.~Aaij {\it et al.} [LHCb Collab.],
%``Angular analysis and differential branching fraction of the decay $B^0_s\to\phi\mu^+\mu^-$,''
JHEP {\bf 1509}, 179 (2015)
%  doi:10.1007/JHEP09(2015)179
%[arXiv:1506.08777 [hep-ex]].
%%CITATION = doi:10.1007/JHEP09(2015)179;%%
%93 citations counted in INSPIRE as of 23 May 2017

%\cite{Altmannshofer:2014rta}
\bibitem{Altmannshofer:2014rta} 
W.~Altmannshofer and D.M.~Straub,
%``New physics in $b\rightarrow s$ transitions after LHC run 1,''
Eur.\ Phys.\ J.\ C {\bf 75}, 382 (2015)
%  doi:10.1140/epjc/s10052-015-3602-7
%[arXiv:1411.3161 [hep-ph]].
%%CITATION = doi:10.1140/epjc/s10052-015-3602-7;%%
%201 citations counted in INSPIRE as of 23 May 2017

%\cite{Straub:2015ica}
\bibitem{Straub:2015ica} 
A.~Bharucha, D.M.~Straub and R.~Zwicky,
%``$B\to V\ell^+\ell^-$ in the Standard Model from light-cone sum rules,''
JHEP {\bf 1608}, 098 (2016)
%  doi:10.1007/JHEP08(2016)098
%[arXiv:1503.05534 [hep-ph]].
%%CITATION = doi:10.1007/JHEP08(2016)098;%%
%118 citations counted in INSPIRE as of 23 May 2017

 %\cite{CMS:2014xfa}
\bibitem{Bsmumu} 
%\cite{Aaij:2017vad}
% \bibitem{Aaij:2017vad} 
R.~Aaij {\it et al.} [LHCb Collab.],
%``Measurement of the $B^0_s\to\mu^+\mu^-$ branching fraction and effective lifetime and search for $B^0\to\mu^+\mu^-$ decays,''
Phys.\ Rev.\ Lett.\  {\bf 118}, 191801 (2017)
%   doi:10.1103/PhysRevLett.118.191801
%[arXiv:1703.05747 [hep-ex]].
%%CITATION = doi:10.1103/PhysRevLett.118.191801;%%
%19 citations counted in INSPIRE as of 25 May 2017
%   V.~Khachatryan {\it et al.} [CMS and LHCb Collaborations],
%   %``Observation of the rare $B^0_s\to\mu^+\mu^-$ decay from the combined analysis of CMS and LHCb data,''
%   Nature {\bf 522}, 68 (2015)
%%   doi:10.1038/nature14474
%   [arXiv:1411.4413 [hep-ex]].
%%CITATION = doi:10.1038/nature14474;%%
%293 citations counted in INSPIRE as of 23 May 2017


%\cite{Bobeth:2013uxa}
\bibitem{BsmumuSM} 
  C.~Bobeth {\it et al.}, 
% M.~Gorbahn, T.~Hermann, M.~Misiak, E.~Stamou and M.~Steinhauser,
%``$B_{s,d} \to l^+ l^-$ in the Standard Model with Reduced Theoretical Uncertainty,''
Phys.\ Rev.\ Lett.\  {\bf 112}, 101801 (2014)
%  doi:10.1103/PhysRevLett.112.101801
%[arXiv:1311.0903 [hep-ph]].
%%CITATION = doi:10.1103/PhysRevLett.112.101801;%%
%229 citations counted in INSPIRE as of 23 May 2017

%\cite{Fleischer:2014jaa}
\bibitem{Fleischer:2014jaa} 
R.~Fleischer,
%``Probing new physics with $B^{0}_s\to\mu^+\mu^-$: Status and perspectives,''
Int.\ J.\ Mod.\ Phys.\ A {\bf 29}, 1444004 (2014);
%doi:10.1142/S0217751X14440047
%[arXiv:1407.0916 [hep-ph]].
%%CITATION = doi:10.1142/S0217751X14440047;%%
%2 citations counted in INSPIRE as of 19 Jun 2017




\bibitem{oldlit}
C.~Bobeth {\it et al.}, % T.~Ewerth, F.~Kruger and J.~Urban,
  %``Analysis of neutral Higgs boson contributions to the decays $\bar{B}$( $s^{)} \to \ell^{+} \ell^{-}$ and $\bar{B} \to K \ell^{+} \ell^{-}$,''
  Phys.\ Rev.\ D {\bf 64}, 074014 (2001);
  %doi:10.1103/PhysRevD.64.074014
%  [hep-ph/0104284];\\
  %%CITATION = doi:10.1103/PhysRevD.64.074014;%%
%  C.~Bobeth, G.~Hiller and G.~Piranishvili,
%  %``Angular distributions of $\bar{B} \to \bar{K} \ell^+\ell^-$ decays,''
%  JHEP {\bf 0712}, 040 (2007)
%  %doi:10.1088/1126-6708/2007/12/040
%  [arXiv:0709.4174 [hep-ph]];\\
  %%CITATION = doi:10.1088/1126-6708/2007/12/040;%%  
   G.~Hiller and M.~Schmaltz,
  %``$R_K$ and future $b \to s \ell \ell$ physics beyond the standard model opportunities,''
  Phys.\ Rev.\ D {\bf 90}, 054014 (2014);
  %doi:10.1103/PhysRevD.90.054014
%  [arXiv:1408.1627 [hep-ph]];\\
  %%CITATION = doi:10.1103/PhysRevD.90.054014;%%  
%
  F.~Beaujean, C.~Bobeth and S.~Jahn,
  %``Constraints on tensor and scalar couplings from $B\rightarrow K\bar{\mu }\mu $ and $B_s\rightarrow \bar{\mu }\mu $,''
  Eur.\ Phys.\ J.\ C {\bf 75}, 456 (2015);
  %doi:10.1140/epjc/s10052-015-3676-2
%  [arXiv:1508.01526 [hep-ph]];\\
  %%CITATION = doi:10.1140/epjc/s10052-015-3676-2;%%
  L.~Calibbi, A.~Crivellin and T.~Ota,
  %``Effective Field Theory Approach to b→sℓℓ(′), B→K(*)ν$\overline{ν}$ and B→D(*)τν with Third Generation Couplings,''
  Phys.\ Rev.\ Lett.\  {\bf 115}, 181801 (2015);
%  doi:10.1103/PhysRevLett.115.181801
%  [arXiv:1506.02661 [hep-ph]].
  %%CITATION = doi:10.1103/PhysRevLett.115.181801;%%
  %112 citations counted in INSPIRE as of 30 Jun 2017
%\cite{Crivellin:2015era}
  A.~Crivellin {\em et al.}, %, L.~Hofer, J.~Matias, U.~Nierste, S.~Pokorski and J.~Rosiek,
  %``Lepton-flavour violating $B$ decays in generic $Z'$ models,''
  Phys.\ Rev.\ D {\bf 92}, 054013 (2015);
%  doi:10.1103/PhysRevD.92.054013
%  [arXiv:1504.07928 [hep-ph]].
  %%CITATION = doi:10.1103/PhysRevD.92.054013;%%
  %78 citations counted in INSPIRE as of 30 Jun 2017
  D.~Be\v{c}irevi\'c  {\em et al.}, 
% D.~Becirevic, S.~Fajfer, N.~Kosnik and O.~Sumensari,
  %``Leptoquark model to explain the $B$-physics anomalies, $R_K$ and $R_D$,''
  Phys.\ Rev.\ D {\bf 94}, 115021 (2016);
  %doi:10.1103/PhysRevD.94.115021
%  [arXiv:1608.08501 [hep-ph]];\\
  %%CITATION = doi:10.1103/PhysRevD.94.115021;%%  
   D.~Das {\it et al.}, % C.~Hati, G.~Kumar and N.~Mahajan,
  %``Towards a unified explanation of $R_{D^{(\ast)}}$, $R_{K}$ and $(g-2)_{\mu}$ anomalies in a left-right model with leptoquarks,''
  Phys.\ Rev.\ D {\bf 94}, 055034 (2016);
%``Scrutinizing $R$-parity violating interactions in light of $R_{K^{(\ast)}}$ data,''
  arXiv:1705.09188 [hep-ph];
  %%CITATION = ARXIV:1705.09188;%%
 %\cite{Choudhury:2016ulr}
% \bibitem{Choudhury:2016ulr} 
  D.~Choudhury {\it et al.}, % A.~Kundu, S.~Nandi and S.~K.~Patra,
 %``Unified resolution of the $R(D)$ and $R(D^*)$ anomalies and the lepton flavor violating decay $h\to\mu\tau$,''
 Phys.\ Rev.\ D {\bf 95}, 035021 (2017);
% doi:10.1103/PhysRevD.95.035021
% [arXiv:1612.03517 [hep-ph]].
 %%CITATION = doi:10.1103/PhysRevD.95.035021;%%
 %4 citations counted in INSPIRE as of 27 May 2017
%  [arXiv:1605.06313 [hep-ph]]; \\
  %%CITATION = doi:10.1103/PhysRevD.94.055034;%%
  %34 citations counted in INSPIRE as of 29 May 2017
%\bibitem{Bhattacharya:2016zcw} 
  S.~Bhattacharya, S.~Nandi and S.K.~Patra,
  %``Looking for possible new physics in $B\to D^{(\ast)}\tau\nu_{\tau}$ in light of recent data,''
  Phys.\ Rev.\ D {\bf 95}, 075012 (2017);
%  doi:10.1103/PhysRevD.95.075012
%  [arXiv:1611.04605 [hep-ph]].
  %%CITATION = doi:10.1103/PhysRevD.95.075012;%%
  %11 citations counted in INSPIRE as of 27 Jun 2017
 B.~Bhattacharya {\it et al.}, 
% A.~Datta, J.~P.~Gu\'evin, D.~London and R.~Watanabe,
  %``Simultaneous Explanation of the $R_K$ and $R_{D^{(*)}}$ Puzzles: a Model Analysis,''
  JHEP {\bf 1701}, 015 (2017);
  %doi:10.1007/JHEP01(2017)015
%  [arXiv:1609.09078 [hep-ph]];\\
  %%CITATION = doi:10.1007/JHEP01(2017)015;%%  
  D.~Bardhan, P.~Byakti and D.~Ghosh,
 %``A closer look at the R$_{D}$ and R$_{D^*}$ anomalies,''
 JHEP {\bf 1701}, 125 (2017);
% [arXiv:1610.03038 [hep-ph]]; \\
  %%CITATION = doi:10.1007/JHEP01(2017)125;%%
   % 
  W.~Altmannshofer, P.S.B.~Dev and A.~Soni,
    %``$R_{D^{(*)}}$ anomaly: A possible hint for natural supersymmetry with $R$-parity violation,''
    arXiv:1704.06659 [hep-ph].
    %%CITATION = ARXIV:1704.06659;%%

%
\bibitem{Altmannshofer:2008dz}
  W.~Altmannshofer {\it et al.}, % P.~Ball, A.~Bharucha {\it et al.},
%``Symmetries and Asymmetries of $B \to K^* \mu^+ \mu^-$ Decays in the Standard 
%Model and Beyond,''
JHEP {\bf 0901}, 019 (2009).
%[arXiv:0811.1214 [hep-ph]].  

\bibitem{rknew}
B.~Capdevila {\it et al.}, 
% A.~Crivellin, S.~Descotes-Genon, J.~Matias and J.~Virto,
  %``Patterns of New Physics in $b\to s\ell^+\ell^-$ transitions in the light of recent data,''
  arXiv:1704.05340 [hep-ph];
  %%CITATION = ARXIV:1704.05340;%%
  W.~Altmannshofer, P.~Stangl and D.M.~Straub,
  %``Interpreting Hints for Lepton Flavor Universality Violation,''
  arXiv:1704.05435 [hep-ph];
  %%CITATION = ARXIV:1704.05435;%%  
  G.~D'Amico {\it et al.}, 
% M.~Nardecchia, P.~Panci, F.~Sannino, A.~Strumia, R.~Torre and A.~Urbano,
  %``Flavour anomalies after the $R_{K^*}$ measurement,''
  arXiv:1704.05438 [hep-ph];
  %%CITATION = ARXIV:1704.05438;%%  
  G.~Hiller and I.~Nisandzic,
  %``$R_K$ and $R_{K^{\ast}}$ beyond the Standard Model,''
  arXiv:1704.05444 [hep-ph];
  %%CITATION = ARXIV:1704.05444;%%
L.S.~Geng {\it et al.}, 
 %B.~Grinstein, S.~J\"ager, J.~Martin Camalich, X.~L.~Ren and R.~X.~Shi,
  %``Towards the discovery of new physics with lepton-universality ratios of $b\to s\ell\ell$ decays,''
  arXiv:1704.05446 [hep-ph];
  %%CITATION = ARXIV:1704.05446;%%
  M.~Ciuchini {\it et al.}, 
% A.~M.~Coutinho, M.~Fedele, E.~Franco, A.~Paul, L.~Silvestrini and M.~Valli,
  %``On Flavourful Easter eggs for New Physics hunger and Lepton Flavour Universality violation,''
  arXiv:1704.05447 [hep-ph];
  %%CITATION = ARXIV:1704.05447;%%
A.~Celis {\it et al.}, % J.~Fuentes-Martin, A.~Vicente and J.~Virto,
  %``Gauge-invariant implications of the LHCb measurements on Lepton-Flavour Non-Universality,''
  arXiv:1704.05672 [hep-ph];
  %%CITATION = ARXIV:1704.05672;%%
D.~Be\v{c}irevi\'c and O.~Sumensari,
  %``A leptoquark model to accommodate $R_K^\mathrm{exp}$",   
  arXiv:1704.05835 [hep-ph];
  %%CITATION = ARXIV:1704.05835;%%
Y.~Cai {\it et al.}, % J.~Gargalionis, M.~A.~Schmidt and R.~R.~Volkas,
  %``Reconsidering the One Leptoquark solution: flavor anomalies and neutrino mass,''
  arXiv:1704.05849 [hep-ph];
  %%CITATION = ARXIV:1704.05849;%%
  J.F.~Kamenik, Y.~Soreq and J.~Zupan,
  %``Lepton flavor universality violation without new sources of quark flavor violation,''
  arXiv:1704.06005 [hep-ph];
  %%CITATION = ARXIV:1704.06005;%%  
  F.~Sala and D.M.~Straub,
  %``A New Light Particle in B Decays?,''
  arXiv:1704.06188 [hep-ph];
  %%CITATION = ARXIV:1704.06188;%%
  S.~Di Chiara {\it et al.}, 
% A.~Fowlie, S.~Fraser, C.~Marzo, L.~Marzola, M.~Raidal and C.~Spethmann,
  %``Minimal flavor-changing $Z'$ models and muon $g-2$ after the $R_{K^*}$ measurement,''
  arXiv:1704.06200 [hep-ph];
  %%CITATION = ARXIV:1704.06200;%%  
  D.~Ghosh,
  %``Explaining the $R_K$ and $R_{K^*}$ anomalies,''
  arXiv:1704.06240 [hep-ph];
  %%CITATION = ARXIV:1704.06240;%%  
  A.K.~Alok {\it et al.}, % D.~Kumar, J.~Kumar and R.~Sharma,
  %``Lepton flavor non-universality in the B-sector: a global analyses of various new physics models,''
  arXiv:1704.07347 [hep-ph];
  %%CITATION = ARXIV:1704.07347;%%  
  %\bibitem{Alok:2017sui}
  A.K.~Alok {\it et al.},
  %``New Physics in $b \to s \mu^+ \mu^-$ after the Measurement of $R_{K^*}$,''
  arXiv:1704.07397 [hep-ph];
C.~Bonilla {\it et al.}, % T.~Modak, R.~Srivastava and J.~W.~F.~Valle,
    %``$U(1)_{B_3-3L_\mu}$ gauge symmetry as the simplest description of $b\to s$ anomalies,''
    arXiv:1705.00915 [hep-ph];
    %%CITATION = ARXIV:1705.00915;%% 
  D.~Bardhan, P.~Byakti and D.~Ghosh,
  %``Role of Tensor operators in $R_K$ and $R_{K^*}$,''
  arXiv:1705.09305 [hep-ph];
  %%CITATION = ARXIV:1705.09305;%%
%\cite{Chiang:2017hlj}
%\bibitem{Chiang:2017hlj} 
  C.W.~Chiang {\it et al.}, % X.~G.~He, J.~Tandean and X.~B.~Yuan,
  %``$R_{K^{(*)}}$ and related $b\to s\ell\bar\ell$ anomalies in minimal flavor violation framework with $Z'$ boson,''
  arXiv:1706.02696 [hep-ph].
  %%CITATION = ARXIV:1706.02696;%%
  %4 citations counted in INSPIRE as of 30 Jun 2017



%\cite{Crivellin:2017ecl}
% \bibitem{Crivellin:2017ecl} 
  \bibitem{crivellin-march17}
  A.~Crivellin, D.~M\"uller and T.~Ota,
  %``Simultaneous Explanation of $R(D^{(*)})$ and $b\to s\mu^+\mu^-$: The Last Scalar Leptoquarks Standing,''
  arXiv:1703.09226 [hep-ph];
  %%CITATION = ARXIV:1703.09226;%%
%\cite{Dorsner:2017ufx}
%\bibitem{Dorsner:2017ufx} 
  I.~Dorsner {\it et al.}, % S.~Fajfer, D.~A.~Faroughy and N.~Košnik,
  %``Saga of the two GUT leptoquarks in flavor universality and collider searches,''
  arXiv:1706.07779 [hep-ph].
  %%CITATION = ARXIV:1706.07779;%%


%\bibitem{Na:2015kha}
%  H.~Na {\it et al.} [HPQCD Collab.],
%  %``$B\to D\ell\nu$ form factors at nonzero recoil and extraction of $|V_{cb}|$,''
%  Phys.\ Rev.\ D {\bf 92}, 054510 (2015)
%  [arXiv:1505.03925 [hep-lat]].
%  %%CITATION = doi:10.1103/PhysRevD.92.054510;%%
%  %14 citations counted in INSPIRE as of 06 May 2016


%%\cite{Kamenik:2008tj}
%\bibitem{Kamenik:2008tj}
%  J.~F.~Kamenik and F.~Mescia,
%  %``$B\to D\tau\nu$ Branching Ratios: Opportunity for Lattice QCD and Hadron Colliders,''
%  Phys.\ Rev.\ D {\bf 78}, 014003 (2008)
%  [arXiv:0802.3790 [hep-ph]].
%  %%CITATION = doi:10.1103/PhysRevD.78.014003;%%
%  %109 citations counted in INSPIRE as of 09 Feb 2016
  
%\bibitem{hfag_rd}
%http://www.slac.stanford.edu/xorg/hfag/semi/summer16/html/RDsDsstar/RDRDs.html


 





%\cite{Buras:2014fpa}
\bibitem{MET_SM} 
   A.J.~Buras {\it et al.}, % J.~Girrbach-Noe, C.~Niehoff and D.~M.~Straub,
 %``$ B\to {K}^{\left(\ast \right)}\nu \overline{\nu} $ decays in the Standard Model and beyond,''
JHEP {\bf 1502}, 184 (2015).
%   doi:10.1007/JHEP02(2015)184
%[arXiv:1409.4557 [hep-ph]].
%%CITATION = doi:10.1007/JHEP02(2015)184;%%
%81 citations counted in INSPIRE as of 23 May 2017

\bibitem{MET_SM_old}
%\cite{Misiak:1999yg}
%\bibitem{Misiak:1999yg} 
M.~Misiak and J.~Urban,
%``QCD corrections to FCNC decays mediated by Z penguins and W boxes,''
Phys.\ Lett.\ B {\bf 451}, 161 (1999);
%doi:10.1016/S0370-2693(99)00150-1
%[hep-ph/9901278]; \\
%%CITATION = doi:10.1016/S0370-2693(99)00150-1;%%
%235 citations counted in INSPIRE as of 02 Jun 2017
%
%\cite{Buchalla:1998ba}
%\bibitem{Buchalla:1998ba} 
G.~Buchalla and A.J.~Buras,
%``The rare decays $K\to \pi \nu\bar\nu$, $B \to X \nu\bar\nu$ and $B \to l^+ l^-$: An Update,''
Nucl.\ Phys.\ B {\bf 548}, 309 (1999).
%doi:10.1016/S0550-3213(99)00149-2
%[hep-ph/9901288];\\
%%CITATION = doi:10.1016/S0550-3213(99)00149-2;%%
%361 citations counted in INSPIRE as of 02 Jun 2017

%\cite{Aaij:2017xqt}
\bibitem{Aaij:2017xqt} 
R.~Aaij {\it et al.} [LHCb Collab.],
%``Search for the decays $B_s^0\to\tau^+\tau^-$ and $B^0\to\tau^+\tau^-$,''
arXiv:1703.02508 [hep-ex].
%%CITATION = ARXIV:1703.02508;%%
%7 citations counted in INSPIRE as of 14 Jun 2017


\bibitem{belle17} 
  J.~Grygier {\it et al.} [Belle Collab.],
  %``Search for $B\to h\nu\bar{\nu}$ decays with semileptonic tagging at Belle,''
  arXiv:1702.03224 [hep-ex].
  %%CITATION = ARXIV:1702.03224;%% 
 
\bibitem{ggl}
S.L.~Glashow, D.~Guadagnoli and K.~Lane,
%``Lepton Flavor Violation in $B$ Decays?,''
Phys.\ Rev.\ Lett.\  {\bf 114}, 091801 (2015).
%doi:10.1103/PhysRevLett.114.091801
%[arXiv:1411.0565 [hep-ph]].
%%CITATION = doi:10.1103/PhysRevLett.114.091801;%%




 \bibitem{pdg}
K.A.~Olive {\em et al.} [Particle Data Group Collab.], 
Chin.\ Phys.\ C {\bf 38}, 090001 (2014) and the 2015 update at {\tt http://pdg.lbl.gov}.

\bibitem{paradisi}
F.~Feruglio, P.~Paradisi and A.~Pattori,
  %``On the Importance of Electroweak Corrections for B Anomalies,''
  arXiv:1705.00929 [hep-ph].
 %%CITATION = ARXIV:1705.00929;%%
  
 \end{thebibliography}
\end{document}